\documentclass[reprint,superscriptaddress,10pt,a4paper,preprintnumbers,nofootinbib]{revtex4-1}
\usepackage{amsmath,amssymb,latexsym}
\usepackage{bm}
\usepackage{graphicx}
\usepackage{hepunits}
\usepackage{slashed}
\usepackage[svgnames]{xcolor}
\usepackage[normalem]{ulem}

\usepackage{textcomp}

\begin{document}

\title{Probing the Higgs trilinear self-coupling through Higgs+jet production}

\author{Jun Gao}
\email{jung49@sjtu.edu.cn}
\affiliation{INPAC, Shanghai Key Laboratory for Particle Physics and Cosmology,
School of Physics and Astronomy, Shanghai Jiao Tong University, Shanghai 200240, China}
\affiliation{Key Laboratory for Particle Astrophysics and Cosmology, Shanghai 200240, China
}
\author{Xiao-Min Shen}
\email{xmshen137@sjtu.edu.cn}
\affiliation{INPAC, Shanghai Key Laboratory for Particle Physics and Cosmology,
School of Physics and Astronomy, Shanghai Jiao Tong University, Shanghai 200240, China}
\affiliation{Deutsches Elektronen-Synchrotron DESY, Notkestr. 85, 22607 Hamburg, Germany}
\author{Guoxing Wang}
\email{wangguoxing2015@pku.edu.cn}
\affiliation{Zhejiang Institute of Modern Physics, School of Physics, Zhejiang University, \\
No. 866 Yuhangtang Road, Hangzhou 310058, China}
\affiliation{Institute for Theoretical Physics Amsterdam and Delta Institute for Theoretical Physics, University of Amsterdam, \\
Science Park 904, 1098 XH Amsterdam, Netherlands}
\affiliation{Nikhef, Theory Group, \\
Science Park 105, 1098 XG, Amsterdam, Netherlands}
\author{Li Lin Yang}
\email{yanglilin@zju.edu.cn}
\affiliation{Zhejiang Institute of Modern Physics, School of Physics, Zhejiang University, \\
No. 866 Yuhangtang Road, Hangzhou 310058, China}
\author{Bin Zhou}
\email{zb0429@sjtu.edu.cn}
\affiliation{INPAC, Shanghai Key Laboratory for Particle Physics and Cosmology,
School of Physics and Astronomy, Shanghai Jiao Tong University, Shanghai 200240, China}

\begin{abstract}
We present the calculation of the next-to-leading order (NLO) electroweak (EW) corrections proportional to the Higgs trilinear self-coupling ($\lambda_{HHH}$) for Higgs boson plus one jet production at the Large Hadron Collider (LHC).
We use the method of large top quark mass expansion to tackle the most challenging two-loop virtual amplitude, and apply the Pad\'{e} approximation to extend the region of convergence of the expansion.
We find that the NLO EW corrections is $0.66\%$ for the total cross section. For the invariant mass distribution and Higgs boson transverse momentum distribution, the NLO corrections are almost flat with their values similar in size. Our results can be used to set extra constraints on $\lambda_{HHH}$ at the LHC.

\end{abstract}

\maketitle

\section{Introduction}

After the discovery of the Higgs boson \cite{1207.7214,1207.7235}, accurately measuring its properties including various couplings becomes one of the top priorities of the Large Hadron Collider (LHC).
The most important reason is that the Higgs boson is related to the spontaneous breaking of electroweak symmetry and is believed to be responsible for the masses of all elementary particles.
Also, the Higgs boson may provide the leading portal to possible Hidden sectors beyond the Standard Model (SM).
In particular, the Higgs trilinear self-coupling ($\lambda_{HHH}$) is the key parameter in the Higgs potential.
The precise determination of its value would give us a better chance to understand electroweak symmetry breaking as well as possible new physics (NP) beyond SM.

The $\lambda_{HHH}$ coupling can be measured directly via the double-Higgs boson production.
The very recent observed constraints by direct measurements are $-0.6<\kappa_{\lambda}=\lambda_{HHH}/\lambda_{HHH}^{\rm SM}<6.6$ at $95\%$ confidence level (CL) \cite{ATLAS:2022jtk}, where $ \lambda_{HHH}^{\rm SM}$ is the Higgs trilinear self-coupling in the SM.
However, only including the double-Higgs production is not enough to get precise constraints.
On the one hand, due to the accidental cancellation between the triangle- and box-type Feynman diagrams at LO in the gluon fusion channel,
the total cross section of double Higgs production is heavily suppressed \cite{Glover:1987nx,Plehn:1996wb}.
On the other hand, when using the double-Higgs production to set the constraints, we need strong assumptions on the Higgs coupling modifiers to other SM particles.

Apart from the double-Higgs production processes, $\lambda_{HHH}$ can also appear as loop effects in higher-order electroweak corrections.
The observed constraints on $\lambda_{HHH}$ by using the combined single- and double- Higgs production are $-0.4<\kappa_{\lambda}<6.3$ at $95\%$ CL \cite{ATLAS:2022jtk}, which is better than the constraints from only double-Higgs production.
More importantly, using the single-Higgs production allows us to relax the assumptions on the coupling modifiers to other SM particles, e.g. the coupling modifier between the Higgs boson and the top quark \cite{ATLAS:2022jtk}.
In order to investigate this kind of loop effect, the so-called $C$-parameters are introduced in the literature \cite{Degrassi:2016wml,Maltoni:2017ims}, where different single-Higgs production channels, including VBF, VH, $t\bar{t}H$ and $tHj$ production processes, are analyzed. %
The authors of \cite{Maltoni:2017ims} have found that the differential distributions of those single-Higgs production processes can provide extra sensitivity to determine $\lambda_{HHH}$.
However, the impacts of differential distributions in the
$gg\rightarrow H+j$ process have not been considered in Ref. \cite{Maltoni:2017ims} because  of the highly non-trivial two-loop Feynman integrals.

Recently, a class of two-loop mixed QCD-EW corrections to $gg\rightarrow H + g$
through a loop of light quarks
is investigated \cite{Bonetti:2020hqh}, which doesn't include the contribution from $\lambda_{HHH}$.
In order to include the effect of $\lambda_{HHH}$, we need to consider the contribution from a class of two-loop mixed QCD-EW corrections with a massive top quark loop.
The calculation would be quite challenging, due to the appearance of two-loop Feynman integrals with two different internal masses $m_t$ and $m_H$, and two Mandelstam variables $\hat{s}$ and $\hat{t}$. These integrals are still unknown analytically so far.
In this case, approximations can be used to give reliable predictions in certain kinematic regions. For example, the analytic expressions up to $\mathcal{O}[1/(m_t^2)^{3}]$ based on the method of large top quark mass expansion for the two-loop scattering amplitudes of $H \rightarrow ggg$ and $H \rightarrow q\bar{q}g$ are given in \cite{Gorbahn:2019lwq}, which are used to present the effect of $\kappa_{\lambda}$ corrections on the Higgs boson transverse momentum ($p_T$) distribution in \cite{DiMicco:2019ngk}.

In this work, we focus on the calculation of the $\lambda_{HHH}$ related NLO EW corrections to $pp\rightarrow H+j$ at the LHC and extract the $C$-parameter for the total cross section as well as differential cross sections.
We use the method of large top quark mass expansion to tackle the challenging multi-scale two-loop Feynman integrals, which has been proved to be quite reliable for $H+j$ production \cite{Chen:2016zka,Chen:2021azt} and double-Higgs production below the $2m_t$ region \cite{Grigo:2015dia,Grazzini:2018bsd}.
In order to extend the range of validity of the large top quark mass expansion, we adopt the Pad\'{e} approximation \cite{hep-ph/9403230, hep-ph/9605392, hep-ph/0102266, 1605.01380, 1611.05881}\footnote{ Note that the Pad\'{e} approximation was not used in \cite{Gorbahn:2019lwq}, which makes the $p_T$ distribution given in \cite{DiMicco:2019ngk} only valid in the range $p_T < m_t$.
 In this work, we show that the top quark mass expansion is not convergent even
in the $p_T \lesssim m_t$ region at $\sqrt{s}=13.6$ TeV.}.
This enable us to get reliable predictions in high energy regions, which may be sensitive to NP beyond SM.

This paper is organized as follows. In Sec.~\ref{sec:Methods} we briefly introduce our notations and present the NLO EW corrections to $pp\rightarrow H+j$ related to $\lambda_{HHH}$ at the LHC.
In Sec.~\ref{sec:Numerical results}, we show our numerical results and give the $C$-parameter at the levels of total cross section and differential cross sections. The conclusion comes in Sec.~\ref{sec:Conclusion}. And we leave the expressions of the form factors to Appendix~\ref{appendix:formfactor}.

\section{Methods}
\label{sec:Methods}
\begin{figure}[b!]
   \centering
   \includegraphics[width=0.45\textwidth]{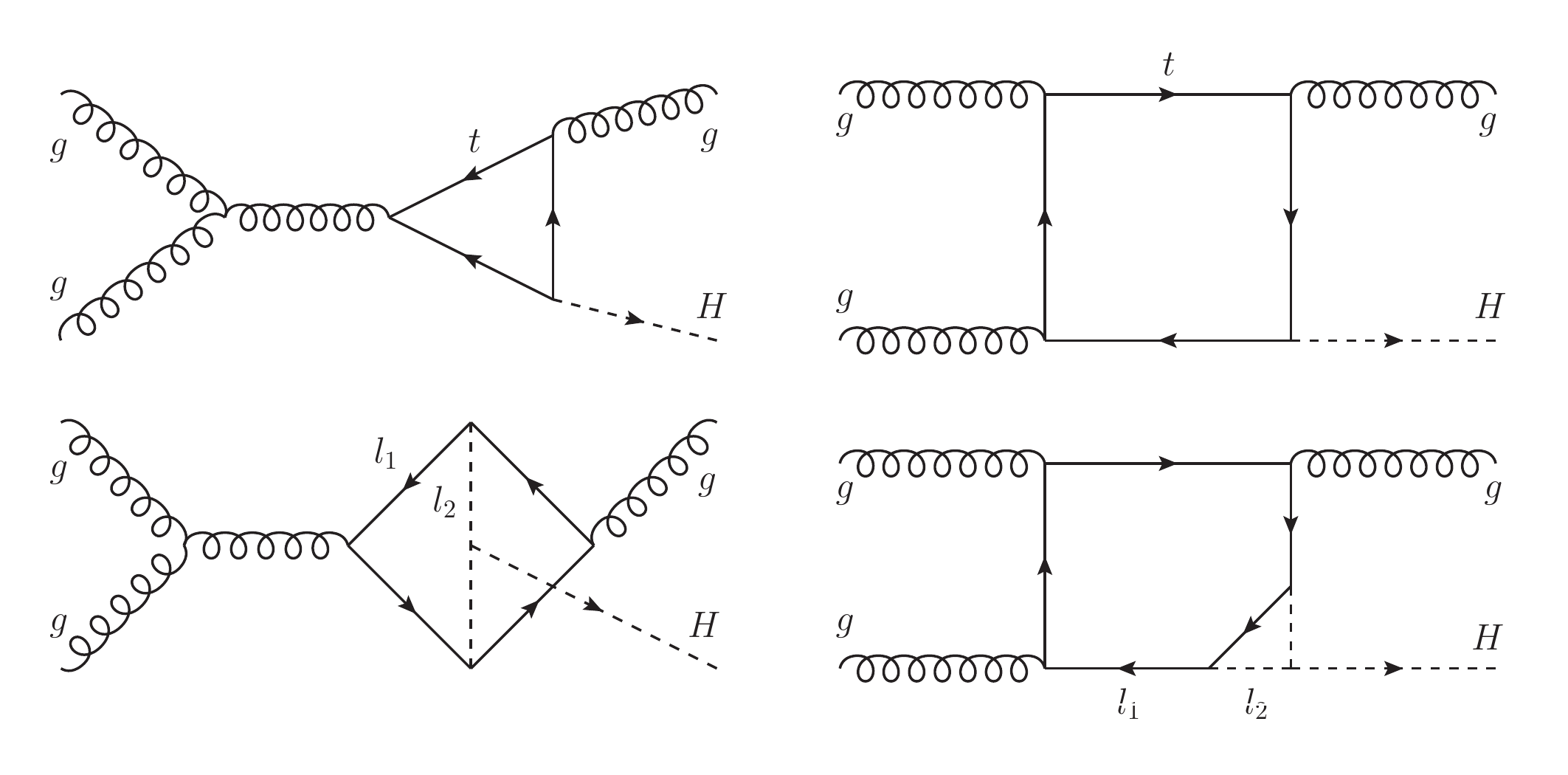}
   \vspace{-3ex}
   \caption{\label{figure:diagram}The typical Feynman diagrams for the gluon fusion channel at LO (upper) and NLO (lower) where dashed lines represent the Higgs bosons, solid lines for the top quarks, and curly lines for the gluons. $l_1$ and $l_2$ are the loop momenta.}
\end{figure}
We need to consider four partonic processes, $g_a(p_1) + g_b(p_2) \to g_c(p_3) + H(p_4)$, $q_a(p_1) + \bar{q}_b(p_2) \to g_c(p_3) + H(p_4)$, $q_a(p_1) + g_b(p_2) \to q_c(p_3) + H(p_4)$ and $\bar{q}_a(p_1) + g_b(p_2) \to \bar{q}_c(p_3) + H(p_4)$,
where $a, b$ and $c$ are color indices, $q(\bar{q})$ only refer to light (anti-)quark,
and $p_i$ are the external momenta with $p_1^2=p_2^2=p_3^2=0$ and $p_4^2=m_H^2$.  We will neglect masses of all
light fermions except that of the top quark in our calculation. Take $gg$ channel as an example, therefore at LO, there are only the diagrams including top quark loop.
Two typical Feynman diagrams at LO are shown in the upper plots of Fig.~\ref{figure:diagram}.
At NLO, we only include the diagrams with a top quark loop and the Higgs trilinear self-coupling $\lambda_{HHH}$ vertex.
With these considerations, there are 21 Feynman diagrams for the gluon fusion channel at NLO and two of them are showed in the lower plots of Fig.~\ref{figure:diagram}.

Before describing the calculation procedures, we first review the definition of the $C$-parameter mentioned in the introduction. Following Ref.~\cite{Degrassi:2016wml}, we consider a beyond SM scenario where the only modification is $\lambda_{HHH}^{\rm SM}$ which can be parameterized via a single parameter $\kappa_{\lambda}$.
Therefore, the Lagrangian containing $\lambda_{HHH}^{\rm SM}$ in the Higgs potential after electroweak-symmetry-breaking can be written as $\kappa_{\lambda}\, \lambda_{HHH}^{\rm SM}\, v \,H^3$, where $v$ is the vacuum expectation value and $H$ is the physical Higgs field.
In presence of the modified trilinear coupling, a generic NLO observable $\Sigma_{\rm NLO}$ (e.g., total or differential cross sections) for single Higgs production can be written as
\begin{align}\label{}
\Sigma_{\rm NLO}=Z_H \Sigma_{\rm LO}(1+ \kappa_{\lambda} C_1)\,,
\end{align}
where $Z_H$ is the Higgs field renormalization constant, $\Sigma_{\rm LO}$ is the LO observable which is not modified with respect to the SM, and $C_1$ is the process- and kinematic-dependent component.
Note that the contribution coming from $Z_H$ is universal and common to all single-Higgs production processes, whose effect can be considered by introducing $C_2$ parameter in \cite{Degrassi:2016wml}. In this paper, we focus on the
process-dependent $C_1$ parameter.
In the limit $ \kappa_{\lambda}\to 1$, $Z_H = 1+ \delta Z_H$, and $\Sigma_{\rm NLO} $ goes to its SM value
\begin{align}\label{eq:sigmaNLO}
\Sigma_{\rm NLO}^{\rm SM}=\Sigma_{\rm LO}(1+  C_1+\delta Z_H)\,,
\end{align}
where we have neglected higher order EW corrections on the right-hand side,
and $\delta Z_H$ is
\begin{align}\label{eq:counterterm}
\delta Z_H = -\frac{9}{16\sqrt{2}\pi^2}\left(\frac{2\pi}{3\sqrt{3}}-1\right)G_Fm_H^2 \,,
\end{align}
with $G_F$ being the Fermi constant. Therefore, $C_1$ can be given by
\begin{align}\label{eq:C1}
C_1&=\frac{\Sigma_{\rm NLO}^{\rm SM}-\Sigma_{\rm LO}-\delta Z_H\,\Sigma_{\rm LO}}{\Sigma_{\rm LO}}\nonumber\\
&=\frac { \sum_{i,j} \int dx_1 dx_2 f_i(x_1) f_j(x_2) \,  2\Re \left (\mathcal{M}_{\rm LO}^{*}\delta\mathcal{M}_{\rm NLO}^{\rm bare} \right) d\Phi_2
 }  { \sum_{i,j}  \int dx_1 dx_2 f_i(x_1) f_j(x_2)  \,  |\mathcal{M}_{\rm LO}|^2  d\Phi_2  }  \, ,
\end{align}
where the sum is over all the possible $i$, $j$ partonic initial states of the process, $f_j(x_k)$ is the parton distribution function (PDF) of the parton $j$ with
a fraction $x_k$ of the initial proton momentum,
$d\Phi_2$ is the two-body phase-space measure,
$\mathcal{M}_{\rm LO}$ is the LO amplitude and
$\delta\mathcal{M}_{\rm NLO}^{\rm bare}$ is the NLO amplitude without  Higgs field renormalization.

We now turn to the calculation of $\mathcal{M}_{\rm LO}$ and $\delta\mathcal{M}_{\rm NLO}^{\rm bare}$.
The amplitudes for the $gg$ and $q\bar{q}$ channels
are given by
\begin{align}\label{eq:amplitude0}
\mathcal{M}^{gg}_{abc} &= \sqrt[4]{2}\sqrt{G_F}\sqrt{4\pi\alpha_s} \, \mathcal{M}_{abc}^{\mu\nu\rho} \epsilon_{\mu}(p_1) \epsilon_{\nu}(p_2) \epsilon^*_{\rho}(p_3)\,, \nonumber \\
\mathcal{M}^{q\bar{q}}_{abc} &=  \sqrt[4]{2}\sqrt{G_F}\sqrt{4\pi\alpha_s} \, \mathcal{M}_{abc}^{\rho}\, \epsilon^*_{\rho}(p_3)\, ,
\end{align}
where $\alpha_s$ is the strong coupling constant. Note that the amplitudes $\mathcal{M}^{qg}_{abc}$ for $qg$ channel and $\mathcal{M}^{\bar{q}g}_{abc}$ for $\bar{q}g$ channel can be obtained by applying the crossing symmetry to $\mathcal{M}^{q\bar{q}}_{abc}$.
$\mathcal{M}_{abc}^{\mu\nu\rho}$ and $\mathcal{M}_{abc}^{\rho}$ can be written as a linear combinations of independent tensor structures respectively:
\begin{align}
  \mathcal{M}_{abc}^{\mu\nu\rho}=& f_{abc} \sum_{i=1}^4 \mathcal{T}_{gg,i}^{\mu\nu\rho}A_{gg,i}(\hat{s}, \hat{t}, m_H, m_t)\,, \label{eq:amplitude1a}\\
  \mathcal{M}_{abc}^{\rho}=& it_{ab}^c \sum_{i=1}^2 \mathcal{T}_{q\bar{q},i}^{\rho}A_{q\bar{q},i}(\hat{s}, \hat{t}, m_H, m_t) \label{eq:amplitude2a}\,,
\end{align}
where the Mandelstam variables are defined as
\begin{align}
&\hat{s} = (p_1+p_2)^2 \, , \quad \hat{t} = (p_1-p_3)^2 \, , \quad \hat{u} = (p_2-p_3)^2 \, ,
\end{align}
which satisfy $\hat{s}+\hat{t}+\hat{u}=m_H^2$.
The four tensor structures in Eq.~\eqref{eq:amplitude1a} are given by
\begin{align}
\mathcal{T}^{\mu\nu\rho}_{gg,1}&=-\frac{1}{2}\hat{s} p_1^{\rho} g^{\mu \nu} +p_2^{\mu } p_1^{\nu } p_1^{\rho }+\frac{\hat{s} \hat{t} p_2^{\rho } g^{\mu \nu }}{2\hat{u}}-\frac{\hat{t} p_2^{\mu } p_1^{\nu } p_2^{\rho}}{\hat{u}}\,,\notag\\
\mathcal{T}^{\mu\nu\rho}_{gg,2}&=   \frac{1}{2} \hat{t} p_1^{\nu } g^{\mu \rho}+p_3^{\mu } p_1^{\nu } p_1^{\rho }+\frac{\hat{s}\hat{t} p_3^{\nu } g^{\mu \rho}}{2\hat{u}}+\frac{\hat{s} p_3^{\mu } p_3^{\nu } p_1^{\rho}}{\hat{u}}\,,\notag\\
\mathcal{T}^{\mu\nu\rho}_{gg,3}&=   \frac{1}{2} \hat{u} p_2^{\mu
   } g^{\nu \rho }+p_2^{\mu } p_3^{\nu } p_2^{\rho }+\frac{\hat{s}
   p_3^{\mu } p_3^{\nu } p_2^{\rho }}{\hat{t}}+\frac{\hat{s} \hat{u} p_3^{\mu } g^{\nu \rho }}{2 \hat{t}}\,,\notag\\
\mathcal{T}^{\mu\nu\rho}_{gg,4}&=   -\frac{1}{2} \hat{s} p_3^{\mu } g^{\nu \rho }+\frac{1}{2} \hat{s} p_3^{\nu} g^{\mu \rho }-\frac{1}{2} \hat{t} p_2^{\mu } g^{\nu \rho}
+\frac{1}{2} \hat{t} p_2^{\rho } g^{\mu \nu }\notag\\
&+\frac{1}{2} \hat{u}
   p_1^{\nu } g^{\mu \rho }-\frac{1}{2} \hat{u} p_1^{\rho } g^{\mu
   \nu }-p_2^{\mu } p_3^{\nu } p_1^{\rho }+p_3^{\mu }
   p_1^{\nu } p_2^{\rho } \,.
\label{eq:tensor1}
\end{align}
And the two tensor structures in Eq.~\eqref{eq:amplitude2a} are given by
\begin{align}
\mathcal{T}^{\rho}_{q\bar{q},1}&=\left[\frac{\hat{t}}{2}\,\bar{v}(p_2)\gamma^{\rho}u(p_1) + \bar{v}(p_2)\slashed{p}_3u(p_1)\,p_1^{\rho}\right]\,,\notag\\
\mathcal{T}^{\rho}_{q\bar{q},2}&=\left[\frac{\hat{u}}{2}\,\bar{v}(p_2)\gamma^{\rho}u(p_1) + \bar{v}(p_2)\slashed{p}_3u(p_1)\,p_2^{\rho}\right]\,.
\label{eq:tensor2}
\end{align}
Note that these tensor structures are organized such that the form factors $A_{gg,i}$ and $A_{q\bar{q},i}$ are gauge invariant.
To calculate the form factors, we generate the relevant Feynman diagrams using \texttt{FeynArts} \cite{Hahn:2000kx}. The resulting amplitudes are further manipulated with \texttt{FeynCalc} \cite{Mertig:1990an,Shtabovenko:2016sxi,Shtabovenko:2020gxv}. Finally, two sets of projection operators constructed from Eq.~\eqref{eq:tensor1} and  Eq.~\eqref{eq:tensor2} are used to extract $A_{gg,i}$ and $A_{q\bar{q},i}$ from the amplitudes $\mathcal{M}_{abc}^{\mu\nu\rho}$ and $\mathcal{M}_{abc}^{\rho}$ respectively, which can be found in Ref.~\cite{Gehrmann:2011aa}. The form factors can be perturbatively expanded according to
\begin{align}
A_{gg\,(q\bar{q}),i} = \frac{\alpha_s}{4\pi}\left[A_{gg\,(q\bar{q}),i}^{(0)} + \frac{G_F}{2\sqrt{2}\pi^2}A_{gg\,(q\bar{q}),i}^{(1)} + \mathcal{O}\left(G_F^2\right)\right]\,,
\end{align}
where $A_{gg\,(q\bar{q}),i}^{(0)}$ are the LO contributions and $A_{gg\,(q\bar{q}),i}^{(1)}$ are the NLO contributions which involve the two-loop Feynman integrals with four independent scales $\hat{s}, \hat{t}, m^2_t$ and $m_H^2$.
For $A_{gg\,(q\bar{q}),i}^{(1)}$, we only calculate the contributions coming from the two-loop Feynman diagrams with a $\lambda_{HHH}$ vertex, denoted by $A_{gg\,(q\bar{q}),i}^{(1),\rm bare}$.
A straightforward calculation of these two-loop Feynman diagrams is very challenging for two reasons.
Firstly, it is very difficult to perform Integration-by-Parts (IBP) reduction for the non-planar integral family.
Secondly, the analytic results of relevant master integrals are by far unknown.
In order to tackle these challenging calculations, we apply the large top quark mass expansion to $A_{gg\,(q\bar{q}),i}^{(1),\rm bare}$.
Based on the method of expansion by regions,
the integration domains of loop momenta $(l_1,l_2)$ are divided into four regions: hard-hard, hard-soft, soft-hard and soft-soft.
According to the definition of $l_1$ and $l_2$ shown in Fig.~\ref{figure:diagram}, only hard-hard and hard-soft regions contribute.
Combining the contribution from these two regions, we obtain the final results of the Feynman integrals
in the limit $m_t^2 \gg \hat{s}, |\hat{t}|, m_H^2$.
In our work, $A_{gg\,(q\bar{q}),i}^{(1),\rm bare}$ are expanded up to $\mathcal{O}[1/(m_t^2)^{6}]$ (N$^6$LP).
The most time consuming part is the IBP reduction of huge number of Feynman integrals after expansion by regions.
At N$^6$LP, there are about 5 million Feynman integrals before IBP reduction.
In the large top quark mass limit, the structures of the form factors $A_{gg\,(q\bar{q}),i}^{(1),\rm bare}$ are very simple.
Schematically, we present the $\mathcal{O}[1/(m_t^2)^{0}]$ (LP) contributions of the $A_{gg,i}^{(1),\rm bare}$.
\begin{align}
\vec{A}_{gg}^{(1),\rm bare}&= \frac{m_H^2}{12} \left(-12 L_m + 4 \sqrt{3} \pi -23\right)\notag\\
&\times\left(\frac{1}{\hat{t}},\frac{1}{\hat{s}},-\frac{1}{\hat{s}},\frac{1}{\hat{s}}+\frac{1}{\hat{t}}+\frac{1}{\hat{u}}\right)\,,
\end{align}
where $L_m = \ln(m_t^2/m_H^2)$ and $\sqrt{3}\pi$ terms come from the hard-soft region of the two-loop integrals and $A_{gg,i}^{(1),\rm bare}$ is the $i$-th element of $\vec{A}_{gg}^{(1),\rm bare}$.
Note that, there are no ultraviolet (UV) and infrared (IR) divergences in the form factors.

To investigate the validity of large top quark mass expansion, we also expand the LO amplitudes up to $\mathcal{O}[1/(m_t^2)^{12}]$. We will compare the approximate results with the exact results at LO in the next section.
For the exact results at LO, we evaluate the one-loop scalar Feynman integrals using program libraries \texttt{LoopTools} \cite{Hahn:1998yk} and \texttt{QCDLoop} \cite{Carrazza:2016gav}. We have checked that our LO results are in good agreement with those of \texttt{MadGraph5\_aMC@NLO} \cite{Alwall:2014hca}.

At NLO, the form factors $A_{gg\,(q\bar{q}),i}^{(1),\rm bare}$ have been given up to N$^{3}$LP in Ref.~\cite{Gorbahn:2019lwq}. It should be noted that the conventions for the form factors are different between Ref.~\cite{Gorbahn:2019lwq} and our work, because of the different choices of tensor structures and normalization factors. After the necessary transformation to arrive at their convention, we find that our results agree with those given in Ref.~\cite{Gorbahn:2019lwq}. For reference, we show the analytic results of  $A_{gg\,(q\bar{q}),i}^{(1),\rm bare}$ up to N$^4$LP in the Appendix~\ref{appendix:formfactor} and give $A_{gg\,(q\bar{q}),i}^{(1),\rm bare}$ up to N$^6$LP in an electronic file attached to the arXiv submission, together with $A_{gg\,(q\bar{q}),i}^{(0)}$ up to N$^{12}$LP used in our work.

Note that the approximation in large top quark mass limit is valid only for kinematic region where $\hat{s}  < 4 m_t^2$.
In order to get reliable predictions beyond that region, we adopt the Pad\'{e} approximation \cite{hep-ph/9403230, hep-ph/9605392, hep-ph/0102266, 1605.01380, 1611.05881}.
Following Ref. \cite{hep-ph/9403230}, we apply Pad\'{e} approximation to the expansion series in
\begin{equation}
  \label{eq:cfm-map}
w \equiv \frac{1 - \sqrt{1 - s' / (4 m_t^2)}}{1 + \sqrt{1 -s' / (4 m_t^2)}} \,,
\end{equation}
instead of in $1/m_t^2$. Note that $s'=\hat{s}$ for $gg$ channel and $q\bar{q}$ channel, and $s'=\hat{t}$ for $qg$ channel. With the above knowledge, we can give our numerical predictions for $C_1$ parameter defined in Eq.~\eqref{eq:C1}.

\section{Numerical results}
\label{sec:Numerical results}
In this section we present our numerical predictions for the total cross section, the $m_{jh}$ distribution and the $p_T$ distribution, where $m_{jh}=\sqrt{(p_3+p_4)^2}$=$\sqrt{\hat{s}}$ is the invariant mass of the Higgs boson and the final state jet, and $p_T$ is the transverse momentum of the Higgs boson.
We choose the input parameters as
 $m_t=\unit{172.5}{\GeV}$, $m_H=\unit{125.25}{\GeV}$ and $G_F=\unit{$1.1663787 \times 10^{-5}$}{\GeV^{-2}}$ \cite{ParticleDataGroup:2022pth}.
If not stated otherwise we choose $\sqrt{s}=\unit{13.6}{\TeV}$ for the hadronic center-of-mass energy.
Our default PDF set is \texttt{CT18NNLO\_as\_0118} \cite{Hou:2019efy}
for the evaluation of both
LO and NLO cross sections.
For the strong coupling constant we use the value provided by \texttt{CT18NNLO\_as\_0118}: $\alpha_s(m_Z)=0.118$.
The default factorization scale $\mu_f$ and renormalization scale $\mu_r$ are chosen as
$\mu_f=\mu_r=(\sqrt{p_T^2+m_H^2}+p_T)/2$.
Last but not least, we choose the cut $p_T \geq \unit{20}{\GeV}$ to get finite predictions for the total cross section as well as the $m_{jh}$ distribution.

\begin{figure}[h!]
\centering
\raisebox{0.0\height}{\includegraphics[width=8.cm,height=8.0cm]{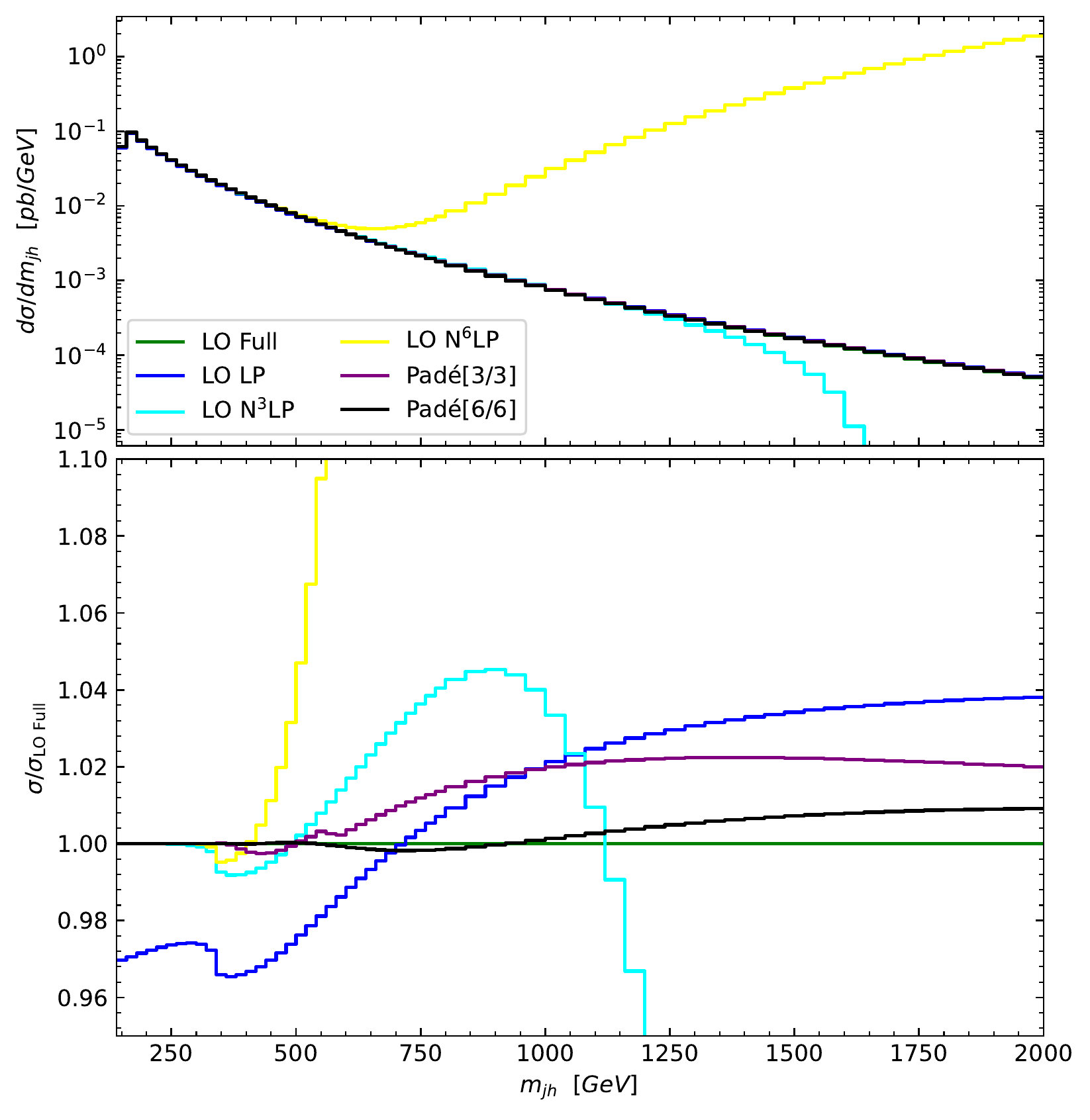}} \\
\caption{The LO differential cross sections of $pp\to H+j$ with respect to the invariant mass $m_{jh}$ of the Higgs boson and the jet. The lower plot shows the ratios to the LO exact values.}
\label{fig:mhj-LO}
\end{figure}

We first present LO results.
We show in Fig.~\ref{fig:mhj-LO} the LO differential cross sections for $m_{jh}$.
In order to see the performance of large top quark mass expansion clearly, we present the exact distribution and the distribution up to $\mathcal{O}[1/(m_t^2)^{n}]$ (N$^{n}$LP).
The upper plot employs a logarithmic scale for the vertical axis to give the distributions in the broad range $\unit{140}{\GeV} \leq m_{jh} \leq \unit{2000}{\GeV}$.
The lower plot shows the ratios to the exact values of the LO differential cross section.
As we expected, the distributions in the region $m_{jh}\leq 2m_t$ show excellent convergence of the large top quark mass expansion.
The relative errors of N$^{3}$LP is already smaller than $0.3\%$.
However, the distributions at N$^{3}$LP and N$^{6}$LP blow up beyond the top quark pair threshold region, which are clear in the lower plot of Fig.~\ref{fig:mhj-LO}.
In order to get reliable estimations in the $m_{jh}\ge 2m_t$ region, we apply the Pad\'{e} approximation to the expansion series in $w$.
Using the expansion series up to N$^{12}$LP, we can construct $[m/n]$ Pad\'{e} approximants with $m+n\leq 12$.
We have shown the distributions with $[3/3]$ and $[6/6]$ Pad\'{e} approximation in  Fig.~\ref{fig:mhj-LO}.
We can find that the relative errors of $[3/3]$ are smaller than $3\%$ and the relative errors of $[6/6]$ are smaller than $1\%$ for $m_{jh} \leq 2000{\GeV}$.

\begin{figure}[h!]
\centering
\raisebox{0.0\height}{\includegraphics[width=8.cm,height=8.0cm]{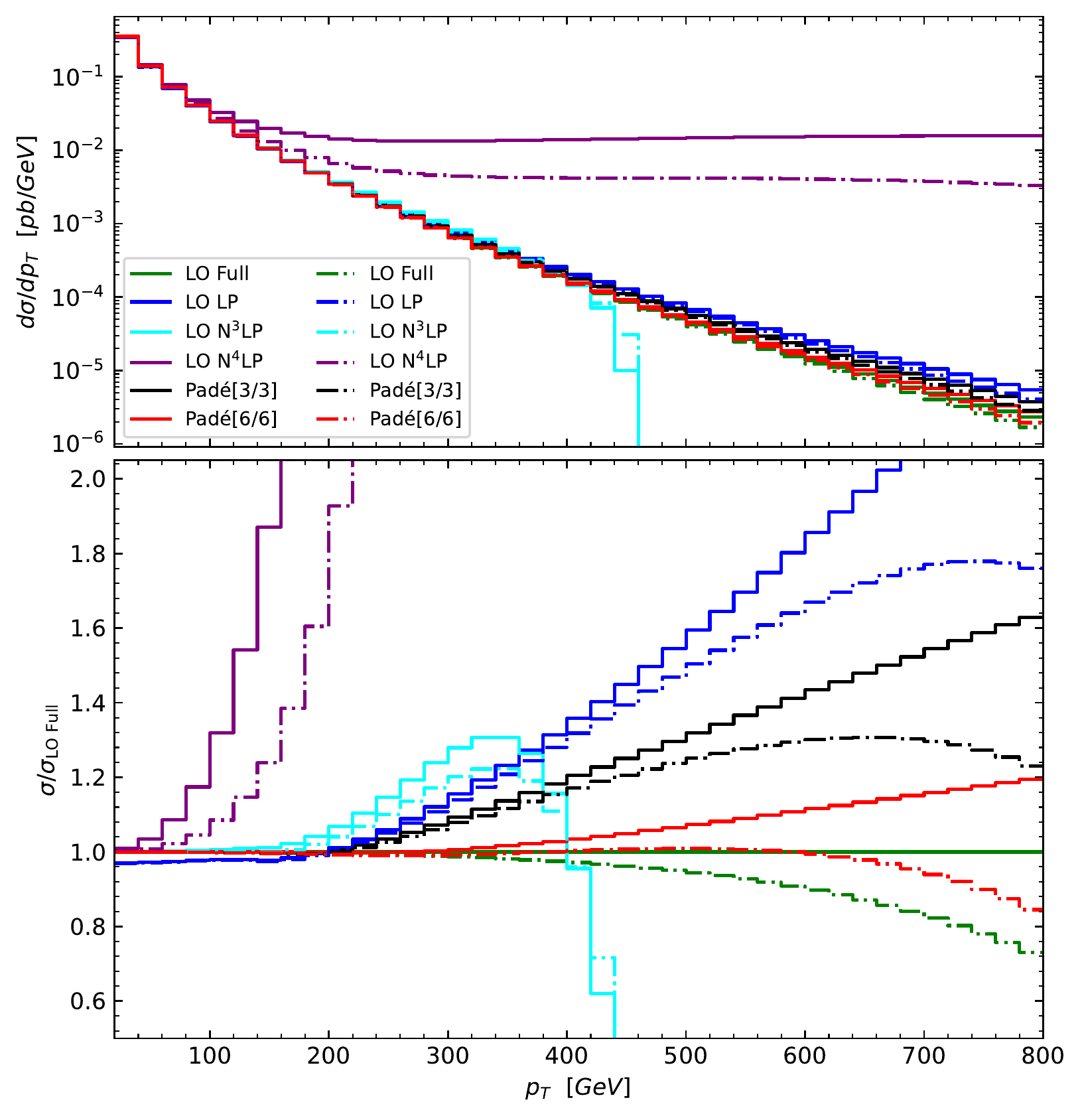}} \\
\caption{The LO differential cross sections of $pp\to H+j$ with respect to the transverse momentum $p_T$ of the Higgs boson. The solid lines show the distributions without the $\hat{s}$-cut, while the dashed lines show the distributions with the $\hat{s}$-cut. The lower plot shows the ratios to the LO exact values without the $\hat{s}$-cut.
}
\label{fig:pt-LO}
\end{figure}

In Fig.~\ref{fig:pt-LO}, we show the $p_T$ distribution at LO.
In the small $p_T$ region, the performance of large top quark mass expansion without/with the Pad\'{e} approximation is very similar to that in $m_{jh}$ distribution.
Both the $[3/3]$ and $[6/6]$ Pad\'{e} approximations may serve as reliable estimations to the exact results.
However, in the large $p_T$ region, the predictions with Pad\'{e} approximation show as much as $20\%$ relative errors, which is significantly different to the case of the $m_{jh}$ distribution.
The Pad\'{e} approximation works better in the large $m_{jh}$ region than in the large $p_T$ region, which can be partly attributed to the following fact. The distribution in the large $p_T$ region only receives the contributions from the large $m_{jh}$ region, while the one in the large $m_{jh}$ region receive dominant contributions from the small $p_T$ region. Because of this, the range of validity for the $p_T$ distribution is much smaller than the naive guess $p_T < m_t$, which can be seen from the lower plot of Fig.~\ref{fig:pt-LO}.
To improve the approximation for the $p_T$ distribution, we impose a novel $\hat{s}$-cut $\sqrt{\hat{s}}\leq\unit{2000}{\GeV}$. The results are shown in Fig.~\ref{fig:pt-LO} as dashed lines.
We find that this $\hat{s}$-cut can improve the convergence to some extent.
For example, the relative errors of [3/3] Pad\'{e} approximation with the $\hat{s}$-cut is less than $1\%$ in the region $p_T\le \unit{240}{\GeV}$ and less than $6\%$ in the region $p_T\le \unit{300}{\GeV}$.
Because the dominant contributions to the total cross section come from the small $m_{jh}$ region, the $\hat{s}$-cut should not give significant influence on the integrated cross section.
We show the LO cross sections integrated over $p_T$ with and without the $\hat{s}$-cut in Table~\ref{tab:cross section}.
We find that the contribution from the region above the $\hat{s}$-cut is only about $0.2\%$. As we expected, the $\hat{s}$-cut makes the approximate results more compatible with the exact one. The $[3/3]$ and $[6/6]$ Pad\'{e} approximations show precise estimations of the exact result.

\begin{table}[h!]
   \centering
   \begin{tabular}{|c|c|c|c|c|c|}
        \hline
        $\hat{s}$-cut &$\sigma_\text{exact}$ &$\sigma_{\text{LP}}$  & $\sigma_{\text{N}^3\text{LP}}$ & $\sigma_{[3/3]}$&$\sigma_{[6/6]}$
        \\ \hline
        no &13.65&13.30&13.08&13.66&13.65
        \\\hline
        yes &13.63&13.28&13.56&13.64&13.63
        \\\hline
   \end{tabular}
   \caption{The LO cross sections (in pb) integrated over $p_T$. $\sigma_{\text{exact}}$ is the exact result. $\sigma_{\text{N}^n\text{LP}}$ are the results with the large top quark mass expansion up to $\mathcal{O}[1/(m_t^2)^{n}]$.  $\sigma_{[n/n]}$ are the results using $[n/n]$ Pad\'{e} approximations.}\label{tab:cross section}
\end{table}

\begin{figure}[h!]
\centering
\raisebox{0.0\height}{\includegraphics[width=8.cm,height=4.5cm]{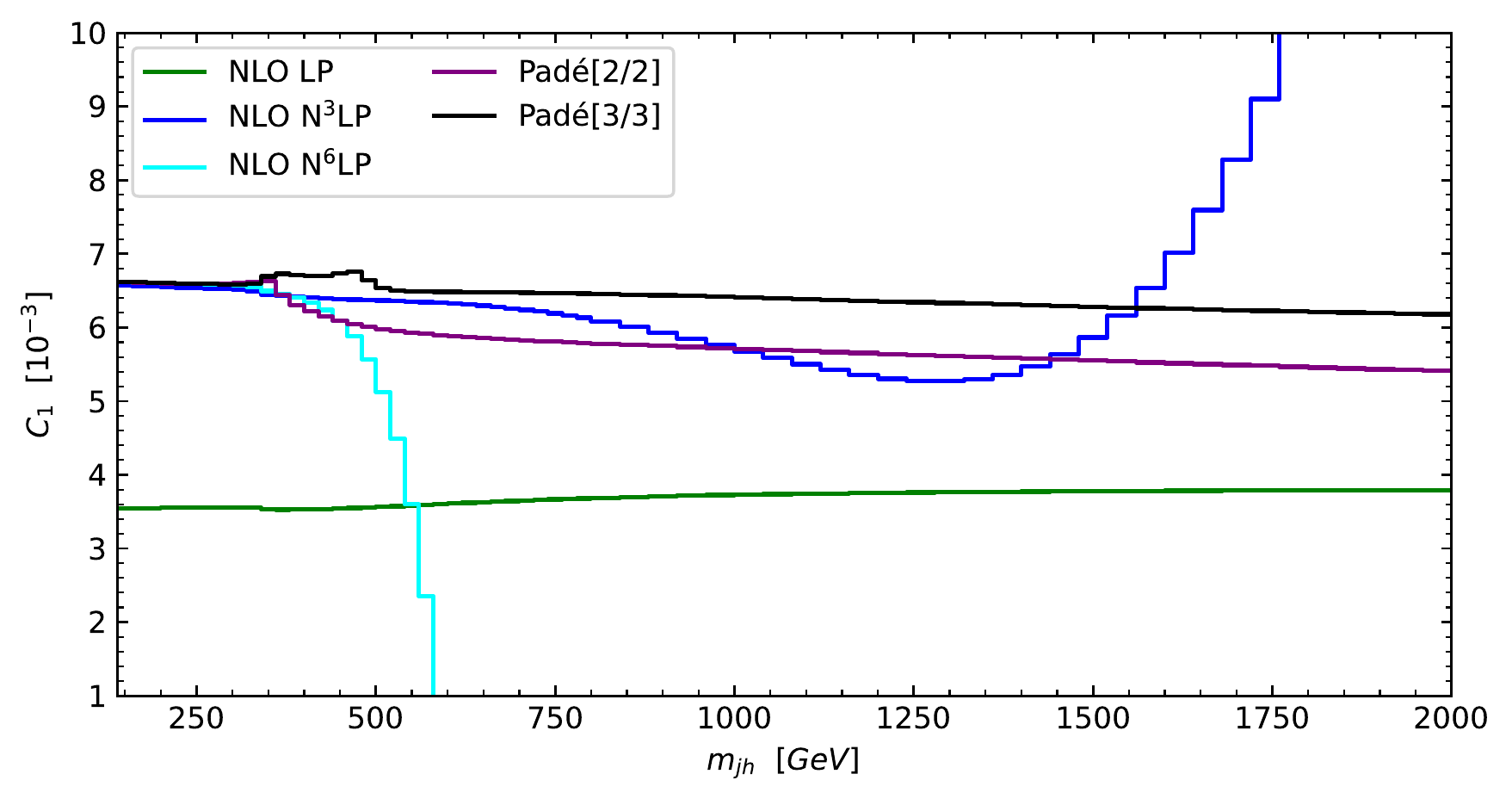}} \\
\caption{
The $C_1$ parameters with respect to the invariant mass $m_{jh}$ of the Higgs boson and the jet.
}
\label{fig:mjh-NLO}
\end{figure}

\begin{figure}[h!]
\centering
\raisebox{0.0\height}{\includegraphics[width=8.cm,height=4.5cm]{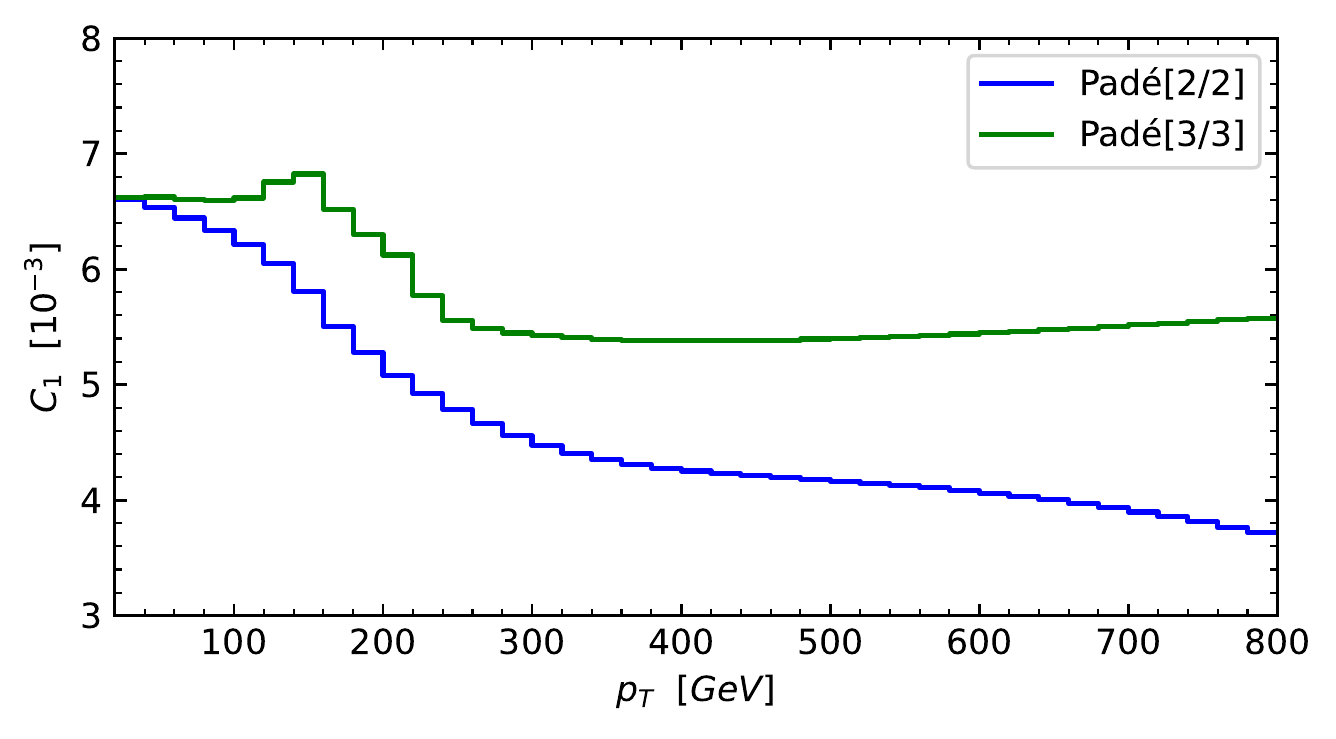}} \\
\caption{
The $C_1$ parameters with respect to the transverse momentum $p_T$ of the Higgs boson. The $\hat{s}$-cut is not applied.
}
\label{fig:pt-NLO}
\end{figure}

We now turn to
give our best predictions of the $C_1$ parameter defined in Eq.~\eqref{eq:C1}.
In Fig.~\ref{fig:mjh-NLO}, we give the $C_1$ parameter in its differential form with respect to $m_{jh}$.
We find that the $\text{N}^6\text{LP}$ and Pad\'{e} approximation give consistent results in the  $m_{jh}\le 2m_t$ region, where the large top quark mass expansion is valid.
These are similar to the cases at LO.
However, there are two significant differences between NLO and LO differential cross sections.
One is that the LP contribution is dominant at LO, while the LP only accounts for about half of the cross section at NLO.
The other is that the trends of $\text{N}^n\text{LP}$(e.g. $n=3$ and $n=6$) curves at NLO are opposite to the corresponding LO ones when $m_{jh} > 2m_t$.
Due to these two differences, the approximation scheme to the full theory (FT$_\text{approx}$) \cite{Maltoni:2014eza}, which is a remarkably reliable method for NLO QCD correction to $pp\to H+j$ \cite{Chen:2016zka,Chen:2021azt}, is not applicable in our case.
Then we also show in Fig.~\ref{fig:pt-NLO} the differential $C_1$ parameter in $p_T$ distribution.
We use the $[3/3]$ Pad\'{e} approximation (without $\hat{s}$-cut in $p_T$ distribution) as our best predictions at NLO.
We note that the differential $C_1$ parameter in both $m_{jh}$ and $p_T$ distributions are almost flat, with its values being
$0.6\%\sim0.7\%$. According to the behavior of the $[3/3]$ Pad\'{e} approximation at LO, we believe that the predictions at NLO are quite
accurate in the convergent region and are reliable in high energy region with their relative errors being at the level of a few percent.
Finally, we give the $C_1$ parameter for total corrections which is defined as the ratio between NLO corrections and LO total cross section
in table~\ref{tab:xsec-NLO}.
Our best prediction at NLO gives $C_1=0.66\%$.
\begin{table}[h!]
   \centering
   \begin{tabular}{|c|c|c|c|c|}
        \hline
        $\hat{s}$-cut &$C_1^\text{LP}$ & $C_1^{\text{N}^3\text{LP}}$ & $C_1^{[2/2]}$&$C_1^{[3/3]}$
        \\ \hline
     no&0.0036& 0.0067& 0.0065& 0.0066\\\hline
     yes&0.0036& 0.0065& 0.0065& 0.0066\\\hline
   \end{tabular}
   \caption{Values of $C_1$ for total corrections.
     $C_1^{\text{N}^n\text{LP}}$ corresponds to large top quark mass expansion up to $\mathcal{O}[1/(m_t^2)^{n}]$.
     $C_1^{[n/n]}$ corresponds to the prediction of $[n/n]$ Pad\'{e} approximation.
     The second line gives the values without the $\hat{s}$-cut, while the third line shows the values with the $\hat{s}$-cut.}\label{tab:xsec-NLO}
\end{table}

\section{Conclusion}
\label{sec:Conclusion}
In this work, we calculate the NLO EW corrections proportional to the Higgs trilinear self-coupling for Higgs boson plus one jet production at the LHC.
We get an asymptotic expansion in the large top quark mass limit with the method of expansion by regions.
The prediction is then extended to high energy regions by applying Pad\'{e} approximation.
We use the $[3/3]$ Pad\'{e} approximation without the $\hat{s}$-cut 
as our best prediction at NLO.
We find the $C_1$ distribution for $m_{jh}$ is rather flat, with its values being $0.6\%\sim0.7\%$ and the behavior of the distribution for $p_T$ is similar.
As for the $C_1$ parameter for total corrections, the value is $0.66\%$.

The high energy behavior of the approximation can be further improved by employing the high energy expansion of \cite{Davies:2018ood,Davies:2018qvx,Mishima:2018olh,Davies:2020drs}, or the small mass expansion of \cite{Xu:2018eos,Wang:2020nnr,Wang:2021rxu}. The latter method is applicable in the whole phase-space region of interest, as was shown in the cases of $gg\to H+H$ and $gg\to H+Z$ processes. Alternatively, one may evaluate the two-loop integrals numerically using the auxiliary mass flow method \cite{Liu:2017jxz,Liu:2020kpc,Liu:2021wks,Liu:2022chg,Liu:2022mfb}. These are left for future investigations.

\begin{acknowledgments}
This work was supported in part by the National Natural Science Foundation of China under Grant No. 11975030, 12147103, 11925506, and the Fundamental Research Funds for the Central Universities.
This work of J. Gao is sponsored by the National Natural Science Foundation of China under the Grant No.12275173 and No.11835005.
The research of G. Wang was supported in part by the International Postdoctoral Exchange Fellowship Program from China Postdoctoral Council under Grant No. PC2021066.
The work of X. Shen is supported in part by the Helmholtz-OCPC Postdoctoral Exchange Program under Grant No. ZD2022004.
\end{acknowledgments}

\appendix
\section{The analytic results for the form factors}\label{appendix:formfactor}
In this appendix, we provide the analytical results for $A_{gg\,(q\bar{q}),i}^{(1),\rm bare}$ up to $\text{N}^4\text{LP}$. First, we present the form factors of the gluon fusion channel. For the $A_{gg,i,\text{NLP}}^{(1),\rm bare}$, we have
\begin{align*}
A_{gg,1,\text{N}\text{LP}}^{(1),\rm bare}&= -\frac{m_H^4}{m_t^2\hat{t}}a_1\,,\notag\\
A_{gg,4,\text{N}\text{LP}}^{(1),\rm bare}&= \frac{m_H^2}{m_t^2\hat{s}}\bigg\{\frac{m_H^2}{\hat{t}\hat{u}}\left[\hat{t}^2+\hat{u}^2-m_H^2\left(\hat{t}+\hat{u}\right)\right]a_1 \,\notag\\
\end{align*}
\vfill
\begin{align}
  + m_H^2a_2 - \left(\hat{t}+\hat{u}\right)a_3 \bigg\} \,,
\end{align}
where
\begin{align}
 a_1&=\frac{7 L_m}{10}-\frac{7 \pi }{20 \sqrt{3}}+\frac{259}{240} \,,\notag \\
 a_2&=\frac{31 L_m}{40}-\frac{17 \pi }{40 \sqrt{3}}+\frac{1487}{1200} \,,\notag \\
 a_3&=\frac{3 L_m}{40}-\frac{\sqrt{3} \pi }{40}+\frac{4}{25} \,.
\end{align}
For the $A_{gg,i,\text{N}^2\text{LP}}^{(1),\rm bare}$, we have
\begin{align}
A_{gg,1,\text{N}^2\text{LP}}^{(1),\rm bare}&= -\frac{m_H^2}{m_t^4\hat{t}}\left[m_H^4b_1 + \hat{t}\hat{u}b_2\right]\,,\notag\\
A_{gg,4,\text{N}^2\text{LP}}^{(1),\rm bare}&= \frac{m_H^4}{m_t^4\hat{s}}\bigg\{\frac{m_H^2}{\hat{t}\hat{u}}\left[\hat{t}^2+\hat{u}^2-m_H^2\left(\hat{t}+\hat{u}\right)\right]b_1 \,\notag\\
&  + m_H^2b_3 - \left(\hat{t}+\hat{u}\right)b_4 \bigg\} \,,
\end{align}
where
\begin{align}
 b_1&=\frac{349 L_m}{1008}-\frac{23 \pi }{240 \sqrt{3}}+\frac{464419}{1058400} \,,\notag \\
 b_2&=\frac{L_m}{140}-\frac{\pi }{280 \sqrt{3}}+\frac{37}{50400}  \,,\notag \\
 b_3&=\frac{1853 L_m}{5040}-\frac{179 \pi }{1680 \sqrt{3}}+\frac{1092101}{2116800} \,,\notag \\
 b_4&=\frac{3 L_m}{140}-\frac{\sqrt{3} \pi }{280}+\frac{54421}{705600} \,.
\end{align}
For the $A_{gg,i,\text{N}^3\text{LP}}^{(1),\rm bare}$, we have
\begin{align}
A_{gg,1,\text{N}^3\text{LP}}^{(1),\rm bare}&= -\frac{m_H^2}{m_t^6\hat{t}}\left[m_H^6c_1 + m_H^2\hat{t}\hat{u}c_2 - \hat{t}\hat{u}\left(\hat{t}+\hat{u}\right)c_3\right]\,,\notag\\
A_{gg,4,\text{N}^3\text{LP}}^{(1),\rm bare}&= \frac{m_H^2}{m_t^6\hat{s}}\bigg\{\frac{m_H^6}{\hat{t}\hat{u}}\left[\hat{t}^2+\hat{u}^2-m_H^2\left(\hat{t}+\hat{u}\right)\right]c_1 \,\notag\\
&  + m_H^6c_4 + m_H^4\left(\hat{t}+\hat{u}\right)c_5  \,\notag\\
&  + \big[\hat{s}\left(2\hat{t}^2+3\hat{t}\hat{u}+2\hat{u}^2\right)+\left(\hat{t}+\hat{u}\right)^3\big]c_6\bigg\} \,,
\end{align}
where
\begin{align}
 c_1&=\frac{1741 L_m}{10800}-\frac{13 \pi }{525 \sqrt{3}}+\frac{31795373}{190512000} \,,\notag \\
 c_2&=\frac{1717 L_m}{126000}-\frac{533 \pi }{126000 \sqrt{3}}+\frac{254311}{79380000} \,,\notag \\
 c_3&=\frac{113 L_m}{42000}-\frac{521 \pi }{126000 \sqrt{3}}+\frac{188969}{45360000} \,,\notag \\
 c_4&=\frac{18541 L_m}{126000}-\frac{407 \pi }{18000 \sqrt{3}}+\frac{9462017}{52920000} \,,\notag \\
 c_5&=\frac{233 L_m}{25200}+\frac{\pi }{360 \sqrt{3}}-\frac{142763}{9072000} \,,\notag \\
 c_6&=\frac{1817 L_m}{378000}-\frac{23 \sqrt{3} \pi }{14000}+\frac{1825337}{476280000} \,.
\end{align}
For the $A_{gg,i,\text{N}^4\text{LP}}^{(1),\rm bare}$, we have
\begin{align}
A_{gg,1,\text{N}^4\text{LP}}^{(1),\rm bare}&= \frac{m_H^2\hat{u}}{m_t^8}\bigg[\frac{m_H^{8}}{\hat{t}\hat{u}}d_1 - m_H^2\left(\hat{t}+\hat{u}\right)d_2+  m_H^4d_3 \notag \\
& + \left(\hat{t}^2+\hat{u}^2\right)d_4+ \hat{t}\hat{u}d_5\bigg] \,,\notag\\
A_{gg,4,\text{N}^4\text{LP}}^{(1),\rm bare}&=\frac{m_H^2}{m_t^8\hat{s}}\bigg\{\frac{m_H^8}{\hat{t}\hat{u}}\left[\hat{t}^2+\hat{u}^2-m_H^2\left(\hat{t}+\hat{u}\right)\right]d_1 \,\notag\\
& + m_H^4\hat{t}\hat{u}d_6+ m_H^4\left(\hat{t}^2+\hat{u}^2\right)d_7+ m_H^8d_8  \, \notag \\
&- \left(\hat{t}+\hat{u}\right)\big[m_H^2\left(\hat{t}^2+\hat{u}^2\right)d_{9} + m_H^2\hat{t}\hat{u}d_{10}\,\notag\\
& - \hat{t}\hat{u}\left(\hat{t}+\hat{u}\right)d_{11} - m_H^6d_{12} \big]\bigg\} \,,
\end{align}
where
\begin{align*}
 d_1&=\frac{10817 L_m}{138600}-\frac{1789 \pi }{277200 \sqrt{3}}+\frac{40370773}{614718720} \,,\notag \\
 d_2&=\frac{55249 L_m}{8316000}-\frac{1067 \pi }{252000 \sqrt{3}}+\frac{106841597}{25613280000} \,,\notag \\
 d_3&=\frac{35473 L_m}{2079000}-\frac{4573 \pi }{1386000 \sqrt{3}}+\frac{240175891}{57629880000} \,,\notag \\
 d_4&=\frac{227 L_m}{1663200}-\frac{19 \pi }{184800 \sqrt{3}}+\frac{272233}{23051952000} \,,\notag \\
 d_5&=\frac{2 L_m}{17325}+\frac{\pi }{11550 \sqrt{3}}-\frac{9083}{34303500} \,,\notag \\
 d_6&=\frac{82219 L_m}{2772000}-\frac{13007 \pi }{924000 \sqrt{3}}+\frac{86605417}{5488560000} \,,\notag \\
 d_7&=\frac{53971 L_m}{2772000}-\frac{26539 \pi }{2772000 \sqrt{3}}+\frac{367056113}{32931360000} \,,\notag \\
 d_8&=\frac{993907 L_m}{16632000}-\frac{29401 \pi }{5544000 \sqrt{3}}+\frac{2544943799}{41912640000} \,,\notag \\
\end{align*}
\vfill
\begin{align}
 d_9&=\frac{53971 L_m}{5544000}-\frac{26539 \pi }{5544000 \sqrt{3}}+\frac{367056113}{65862720000} \,,\notag \\
 d_{10}&=\frac{59021 L_m}{5544000}-\frac{23389 \pi }{5544000 \sqrt{3}}+\frac{243249443}{65862720000} \,,\notag \\
 d_{11}&=\frac{101 L_m}{221760}+\frac{\pi }{3520 \sqrt{3}}-\frac{4126889}{4390848000} \,,\notag \\
 d_{12}&=\frac{7111 L_m}{831600}+\frac{\pi }{275 \sqrt{3}}-\frac{1058129}{1707552000} \,.
\end{align}
Due to the crossing symmetry, we have
\begin{align}
A_{gg,2,\text{N}^j\text{LP}}^{(1),\rm bare} &=A_{gg,1,\text{N}^j\text{LP}}^{(1),\rm bare}(\hat{s}\leftrightarrow \hat{t}) \,,\notag \\
A_{gg,3,\text{N}^j\text{LP}}^{(1),\rm bare}&=-A_{gg,2,\text{N}^j\text{LP}}^{(1),\rm bare}(\hat{t}\leftrightarrow \hat{u})
\end{align}

Then, we give the form factors of the quark-anti-quark annihilation channel. For the $A_{q\bar{q},i,\text{LP}}^{(1),\rm bare}$, we have
\begin{align}
A_{q\bar{q},1,\text{LP}}^{(1),\rm bare}&= \frac{m_H^2}{\hat{s}}\left(\frac{L_m}{2}-\frac{\pi }{2 \sqrt{3}}+\frac{23}{24}\right)\,.
\end{align}
For the $A_{q\bar{q},1,\text{NLP}}^{(1),\rm bare}$, we have
\begin{align}
A_{q\bar{q},1,\text{N}\text{LP}}^{(1),\rm bare}&= \frac{m_H^2}{\hat{s}}\left(m_H^2a^q_1 + \hat{s}a^q_2\right)\,,
\end{align}
where
\begin{align}
 a^q_1&=\frac{7 L_m}{20}-\frac{7 \pi }{40 \sqrt{3}}+\frac{259}{480} \,,\notag \\
 a^q_2&=\frac{11 L_m}{90}-\frac{11 \pi }{120 \sqrt{3}}+\frac{863}{7200} \,.
\end{align}
For the $A_{q\bar{q},1,\text{N}^2\text{LP}}^{(1),\rm bare}$, we have
\begin{align}
A_{q\bar{q},1,\text{N}^2\text{LP}}^{(1),\rm bare}&= \frac{m_H^2}{\hat{s}}\left(m_H^2\hat{s}b^q_1 + m_H^4b^q_2 + \hat{s}^2b^q_3\right)\,,
\end{align}
where
\begin{align}
 b^q_1&= \frac{1271 L_m}{10080}-\frac{167 \pi }{3360 \sqrt{3}}+\frac{266837}{2116800}  \,,\notag\\
 b^q_2&= \frac{349 L_m}{2016}-\frac{23 \pi }{480 \sqrt{3}}+\frac{464419}{2116800}  \,,\notag\\
 b^q_3&= \frac{11 L_m}{420}-\frac{11 \pi }{840 \sqrt{3}}+\frac{17}{5040} \,.
\end{align}
For the $A_{q\bar{q},1,\text{N}^3\text{LP}}^{(1),\rm bare}$, we have
\begin{align}
A_{q\bar{q},1,\text{N}^3\text{LP}}^{(1),\rm bare}&= \frac{m_H^2}{\hat{s}}\left(m_H^6c^q_1 + m_H^4\hat{s}c^q_2 + m_H^2\hat{s}^2c^q_3 + \hat{s}^3c^q_4\right)\,,
\end{align}
where
\begin{align}
 c^q_1&= \frac{1741 L_m}{21600}-\frac{13 \pi }{1050 \sqrt{3}}+\frac{31795373}{381024000} \,,\notag \\
 c^q_2&= \frac{125863 L_m}{1512000}-\frac{9109 \pi }{504000 \sqrt{3}}+\frac{1480511}{19440000} \,,\notag \\
 c^q_3&= \frac{19483 L_m}{504000}-\frac{731 \pi }{72000 \sqrt{3}}+\frac{1296991}{79380000} \,,\notag \\
 c^q_4&= \frac{301 L_m}{54000}-\frac{\sqrt{3} \pi }{1750}-\frac{5466617}{1905120000} \,.
\end{align}
For the $A_{q\bar{q},1,\text{N}^4\text{LP}}^{(1),\rm bare}$, we have
\begin{align}
A_{q\bar{q},1,\text{N}^4\text{LP}}^{(1),\rm bare}&= \frac{m_H^2}{\hat{s}}\big(m_H^8d^q_1 + m_H^6\hat{s}d^q_2 + m_H^4\hat{s}^2d^q_3 \notag \\
 & + m_H^2\hat{s}^3d^q_4 + \hat{s}^4d^q_5\big)\,,
\end{align}
where
\begin{align}
 d^q_1&= \frac{10817 L_m}{277200}-\frac{1789 \pi }{554400 \sqrt{3}}+\frac{40370773}{1229437440} \,,\notag \\
 d^q_2&= \frac{28151 L_m}{594000}-\frac{4999 \pi }{924000 \sqrt{3}}+\frac{17411038151}{461039040000} \,,\notag \\
 d^q_3&= \frac{52319 L_m}{1512000}-\frac{1051 \sqrt{3} \pi }{616000}+\frac{8206234093}{461039040000} \,,\notag \\
 d^q_4&= \frac{184463 L_m}{16632000}-\frac{2813 \pi }{1848000 \sqrt{3}}-\frac{557260999}{461039040000} \,,\notag \\
 d^q_5&= \frac{997 L_m}{831600}-\frac{79 \pi }{277200 \sqrt{3}}-\frac{29741}{28459200} \,.
\end{align}
Due to the crossing symmetry, we have $A_{q\bar{q},2,\text{N}^j\text{LP}}^{(1),\rm bare}=A_{q\bar{q},1,\text{N}^j\text{LP}}^{(1),\rm bare}$.

\bibliographystyle{apsrev4-1}
\bibliography{cite}

\end{document}